\documentclass[fleqn,twoside]{article}
\usepackage{espcrc2}

\usepackage{graphicx}
\usepackage[figuresright]{rotating}

\newcommand{\AmS}{{\protect\the\textfont2
  A\kern-.1667em\lower.5ex\hbox{M}\kern-.125emS}}

\title{
 \begin{flushright}
      {\small TAUP 2764-04\\
      February 15, 2004\\}
   \end{flushright} 
Using a neural network approach for muon reconstruction and triggering}

\author{E. Etzion\address[TAU]{School of Physics and Astronomy, Raymond and Beverly Sackler Faculty of Exact Sciences,\\
Tel Aviv University, 69978 Tel Aviv, Israel \\ }%
        \thanks{Currently at CERN, European Organization for Nuclear Research, EP/ATM, Mailbox E26700, CH-1211, Geneve 23, 
                Switzerland.}
        \thanks{The research was partely supported by the Israeli Science Foundation},
        H. Abramowicz \addressmark, 
        Y. Benhammou\addressmark, 
        G. Dror\address{Department of Computer Sceince, The Academic College of Tel Aviv Yaffo,\\
                         64044 Tel Aviv, Israel\\},
        D. Horn\addressmark[TAU], 
        L. Levinson\address{Department of Particle Physics, Weizmann Institute of Sciences, 76100 Rehovot, Israel}
        and
        R. Livneh\addressmark[TAU]
}       
\begin{document}


\begin{abstract}
{\it A talk given at ACAT03, submitted to Nuclear Instruments and Methods in  Physics Research, Section A \\ \\} 
The extremely high rate of events that will be produced in the future Large Hadron Collider 
requires the triggering mechanism to take precise decisions in a few nano-seconds. We present a study 
which used an artificial neural network  triggering  algorithm and compared it to the  performance of a
dedicated electronic muon triggering system.
Relatively simple architecture was used to solve a complicated inverse problem.
A comparison with a realistic example of the ATLAS first level trigger simulation was in favour of 
the neural network.
A similar architecture trained after the simulation of the electronics first trigger stage showed a further 
background rejection.
\vspace{1pc}
\end{abstract}

\maketitle

\section{INTRODUCTION}
The Large Hadron Collider (LHC) is currently being built at the European
Organization for Nuclear Physics (CERN), Geneva. It will collide beams 
of protons at center of mass energies of $14~TeV$ with a bunch crossing rate 
of $40~MHz$. The design luminosity of the accelerator is 
$10^{34}cm^{-2}s^{-1}$. At this luminosity an average of 23 proton proton 
interactions will be produced at each of the bunch crossing.
A selective fast triggering mechanism is required in order to efficiently 
select the relevant data and eliminate the large amount of background arising 
in these conditions.
The challenge faced at the LHC is to reduce the event rate from an 
interaction rate of about $1~GHz$ by about seven orders of magnitude to an 
event rate of $O(100)~Hz$.
A three layer trigger system was designed for the ATLAS detector\cite{LVL1,HLT} in order to
reach this goal.
The designed trigger system relies on the concept of  
searching for physics objects. 
The selection criteria is mostly based on a single and 
di-object  high $P_T$ trigger, where  high $P_T$ refers to objects such
as charged leptons or jets with large transverse momenta (typically above $O(10)~GeV/C$). 
These objects are crudely reconstructed at the first level of the trigger (LVL1),
where the next stages refine the reconstruction, reject fake selected objects
and improve the selection and reconstruction precision.
We present an artificial neural network (ANN) algorithm which 
is used to trigger events and is  based on the reconstruction of the $P_T$ of muons reaching 
the ATLAS endcap.
We compare the ANN performance to the simulated ATLAS LVL1 endcap muon 
trigger system. A similar ANN is trained on events that pass the 
LVL1 simulation and we show that without additional information the ANN can further 
improve the selection by rejecting fake objects with minimal effect on the acceptance.

\section{THE TRIGGER SYSTEM}
\begin{figure}[htb]
\vspace{9pt}
\includegraphics[scale=0.5]{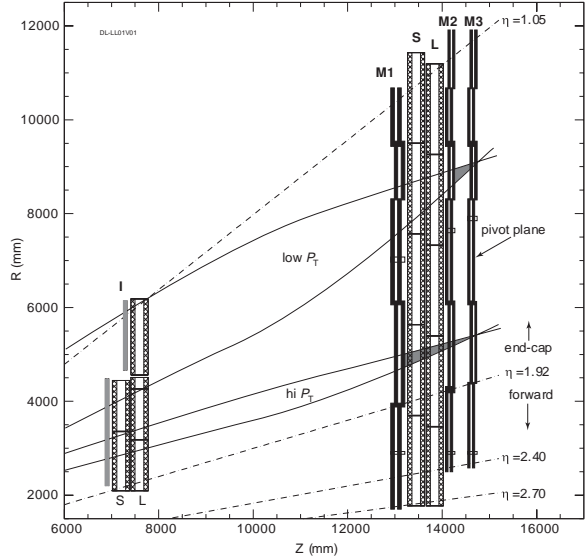}
\caption{Longitudinal view of the TGC system. Low $P_T$ trigger is derived from hits in 'M2' and 'M3' 
doublet layers, where high $P_T$ adds hits in 'M1' triplet layer.}
\label{FIG-TGC}
\end{figure}

The ATLAS experiment LVL1 trigger\cite{LVL1} makes an initial selection based on reduced 
granularity data from a subset of detectors. It uses  calorimetry 
and muon detectors information separately. High $P_T$ muons are identified using dedicated
fast trigger chambers: Resistive Plate Chambers (RPC) in the barrel region and Thin
Gap Chambers (TGC)\cite{ref-TGC} in the endcap region ($1.05 < \eta < 2.7$).
The TGCs are trapezoid shapeed units with area of $1-10~m^2$ arranged in the two
 ATLAS endcaps in four disks at distances
of 7, 13, 14.5 and 15 meters from the interaction point (IP) 
(see figure~\ref{FIG-TGC}). 
The two furthermost disks 'M3' and 'M2' are made of ``doublet units'' which contain two anode
 wire layers laid in the azimuthal direction which provide the 
$r$ coordinate and two orthogonal cathode strips which measure the $\phi$ coordinate. The third disk
'M1' consists of ``triplet units'' which also have two layers of strips but three layers of wires.
The trigger decision is based on the small deflection of high $P_T$ muons. Dedicated electronic
boards calculate the deviation of the muon hits from a straight line drawn between
the IP and the hits in the 'M3' station.
Low $P_T$ trigger is derived between the last two disks 'M3' and 'M2'. For High $P_T$ muon 
the triggers adds the hits information from layer 'M1'. 
As a result of the steeply-falling muon-$P_T$ spectrum, the measured rate of muons  above a given
$P_T$ threshold depends strongly on the reconstruction resolution. 

The material through which a muon passes on its way to the TGCs makes this task more difficult.
The material in front of the trigger chambers is that of the the magnetic toroids, the inner detector
and the hadronic and electromagnetic calorimeters. The total material between the IP and 
TGCs varies between 10 to 15 absorption lengths with somewhat more material in the region of $|\eta| \sim 1.5$.
The ATLAS magnet system\cite{magnet} consists of three air-core superconducting toroids. 
The magnetic field provides typical power of three Tm in the barrel and six Tm in the endcap region.
Owing to the finite number of coils the field configuration is not perfectly toroidal. This inhomogeneity 
which is enhanced in the transition region between the barrel and the endcap, significantly impacts upon the 
resolution of momentum measurements.

\section{THE NEURAL NETWORK}

We present  a selection mechanism which uses an ANN trained to read  the hits in the TGCs and 
reconstruct the transverse momentum of muons. The general structure of the ANN was described elsewhere\cite{ACAT00}.
The network was trained with simulated events generated with the DICE\cite{DICE} simulation program.
The network is a back-propagation feed-forward ANN with two  hidden layers, each of which has ten 'tansig'  
(hyperbolic tangent sigmoid transfer function) neurons. 
One endcap octant in the range of $1.05 <\eta <2.4$ and $0.39< \phi < 1.18$ was divided into 16 regions. 
A different network was trained separately in 
each of these regions. The four input parameters of the network were the muon tracks intercept 
points and the slopes of a linear fit to the hit positions in $xz$ and $yz$ planes.
The output layer has one 'purelin'  (linear transfer function) neuron, set as a  $P_T$ threshold cut.
The simulated sample contained 50,000  events where 1/3 of them were used in a 1,000 epochs training stage.

The simplicity of the ANN structure enabled a fast hardware  realization. This was directly demonstrated by 
the Pierre et Marie Curie University group\cite{DENBY} which created and tested a hardware dedicated 
machinery that successfully implemented 
this ANN structure.

\begin{figure}[htb]
\vspace{9pt}
\includegraphics[scale=0.53]{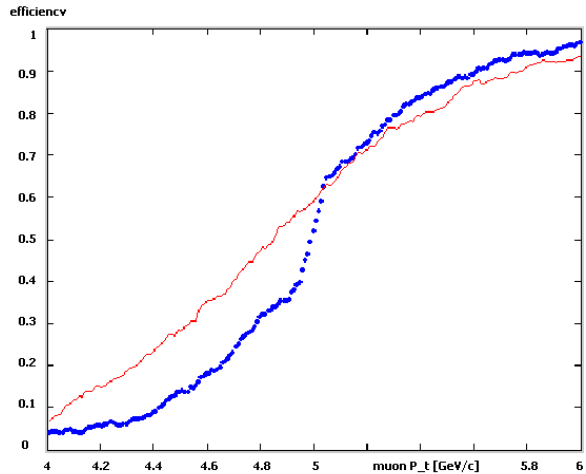}
\caption{A comparison between the simulated electronics selection and the ANN. The $y$ axis is the selection efficiency where the
$x$ axis is the muon's $P_T$. 
The red line is the electronic simulation selection where the blue line is the ANN.}
\label{FIG-ANN-ATR}
\end{figure}

The performance of our network was compared to the results of ATRIG\cite{ATRIG}, a program which  simulates 
the  LVL1 electronics, algorithms and decision menus.
The comparison was done separately in each of the octant 16 regions.
With a similar efficiency in all the regions the ANN selected fewer fake events than the LVL1 simulation.
Figure~\ref{FIG-ANN-ATR} shows a performance summary of all 16 regions. Here it is clearly shown that the 
ANN, which was tuned to select $P_T>5~GeV/C$ events, has a sharper rise around the tuned transverse 
momentum. Namely it selects the signal with slightly higher efficiency than the LVL1 simulation. Furthermore,
 its background rejection is better.

\begin{figure}[htb]
\vspace{8pt}
\includegraphics[scale=0.66]{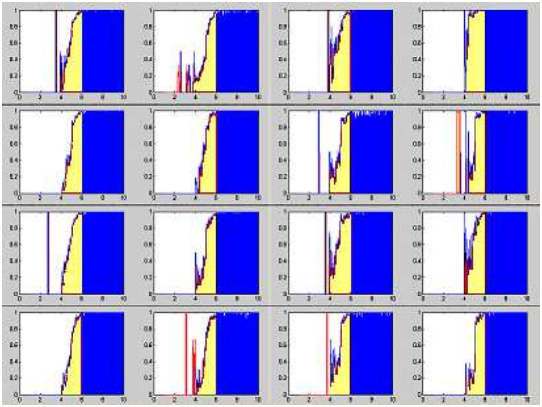}
\caption{The rate of events selected by the ANN applied after LVL1 selection in the 16 regions. The $y$ axis is the efficiency rate, the $x$ axis is the muon's $P_T$. 
The blue regions are the selected events with $P_T>6~GeV/C$, where the yellow ones are 
the rate of fake events selected by the ANN with respect to LVL1 fake events.}
\label{FIG-ANN-LVL2}
\end{figure}

A similar study was conducted training and testing the ANN only on events that passed the LVL1
muon simulation. The ANN was trained to select muons with $P_T>5~GeV/C$. As expected from the first 
study there was only negligible effect on the selection efficiency of LVL1 simulation, 
however the ANN further rejected fake events which were selected by the LVL1 algorithm. 
The results of this study are shown in figure~\ref{FIG-ANN-LVL2} 
for all 16 regions of the tested octant. 
In all 16 regions the ANN rejects almost all the fake events with $P_T\sim 4~GeV/C$, 
about $50\%$ of the events with $P_T\sim 5~GeV/C$ and has negligible imapct on events above the required 
selection threshold.

\begin{figure}[htb]
\vspace{6pt}
\includegraphics[scale=0.66]{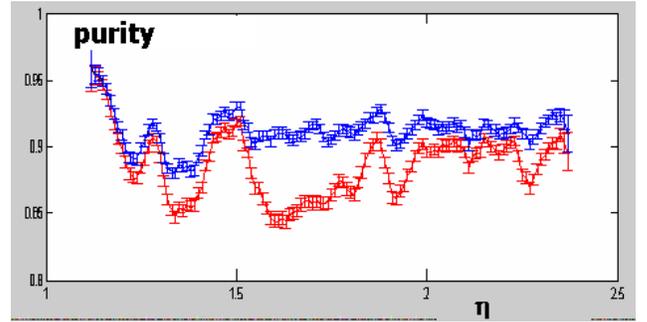}
\caption{The purity of LVL1 simulation (red points) and the ANN (blue points) as a function of $\eta$}
\label{FIG-PUR}
\end{figure}

Figure~\ref{FIG-PUR} shows a comparison between the purity (the ratio between correctly selected events and events
with transverse momentum above the required threshold) of the ANN and the LVL1 simulation as a function 
of the pseudo-rapidity angle, $\eta$.
One can see that both the LVL1 algorithm and the ANN are less sensitive in the region between $1.25<\eta<1.4$ 
where they are affected by the larger quantity of absorbing material before the TGCs and mainly by the inhomogeneous magnetic field in
that area. Nevertheless there are certain regions mainly between $1.5<\eta<1.85$ in which the purity of the 
LVL1 simulation drops close to $85\%$, in comparison to the purity of the network which is mostely stable above $90\%$.

Another attempt was to use the ANN in order to tune the existing electronics coincidence matrix algorithm
to behave similarly to the ANN selection.
It starts with a normal ANN of  the slopes and origins of events
in the $xz$ and $yz$ planes.  From this data we create virtual hits in
two adjacent planes. These are hits a real event might have produced. 
We take the new points on the first plane and shift them in
$x$ and $y$ to the left and right (or up/down). For each shift we create a new
origin/slope in the $xz$ and $yz$ planes. These new virtual trajectories are fed into the 
ANN and the performance of the ANN is plotted with respect to the shift, thus creating a coincidence matrix.

\section{SUMMARY}
A relatively simple feed-forward architecture was used to solve a complicated 
inverse problem of triggering based on the transverse momentum reconstruction.
The simplicity of the network enables very fast hardware realization. 
A similar ANN can be used very efficiently in 
a classification problem necessary for triggering purposes. 
A comparison with a realistic example of first level trigger simulation is in 
favor of the ANN.
A similar architecture trained after simulation of the ATLAS LVL1 electronics 
trigger showed  further  background rejection. 

\end{document}